\documentclass{ws-procs9x6}

\begin{document}

\title{On the Effective Evolution for the Inflaton\footnote{\uppercase{W}ork partially
supported by \uppercase{FAPERJ} and \uppercase{CNP}q.}}

\author{RUDNEI O. RAMOS}

\address{Departamento de F\'{\i}sica Te\'orica,
Universidade do Estado do Rio de Janeiro,\\
20550-013 Rio de Janeiro, RJ, Brazil\\
E-mail: rudnei@uerj.br}

\maketitle

\abstracts{The dynamics of the inflaton field is studied in the context
of its interaction with bosonic and fermionic fields modeled by a
minimal SUSY like model. 
}


Despite the many studies on the dynamics of inflation, much
still remains to be done in the context of understanding several 
aspects concerning the microscopic dynamics underlying many of the
models for inflation, particularly in those cases where the inflaton is
coupled to several other fields, like in hybrid inflation models and
supersymmetric (SUSY) model extensions. In most of these models it is a
fact that large regions of parameter space still remain unexplored and
that can be feasible to inflation and new phenomena. This is also true
when we consider the common approximations used to study the different
aspects of inflation, which most of the time have been
restricted to particular perturbative and linear regimes. This is
understandable since when nonperturbative and nonlinear effects may
become important standard techniques
may not always apply, which
have slowed down progress in that direction. At the same time methods
and techniques developed in quantum field theory devoted to the
description of nonequilibrium dynamics have become essential to study
these new phenomena, like in those cases where there can be
non-negligible particle and radiation production, e.g. in the description
of the preheating phase after isentropic inflation or the study of the
emergence of non-isentropic inflation (or warm inflation) scenarios,
where the scalar inflaton field dissipates non-negligible amounts of
radiation during inflation \cite{bf,hr,bgr,br1}. 

Of special interest as concerned to non-isentropic inflationary scenarios
is the process of how the inflaton can dissipate its energy during
inflation. We have recently identified an efficient mechanism for that
in the context of a nonlinear and nonperturbative regime for the
inflaton dynamics and elaborated on its details in
[~\refcite{br2,br3,br4}]. We consider the
inflaton field $\phi$ in interaction with other scalar and fermion
fields, with standard interaction terms of the form

\begin{equation}
L_{\rm int} =  -\frac{1}{2}g^2 \phi^2 \chi^2 - 
g' \phi {\bar \psi_{\chi}} \psi_{\chi} - h \chi {\bar \psi_d}\psi_d ,
\label{Lint}
\end{equation}

\noindent 
for field masses satisfying $m_\chi > {\rm min}(2
m_{\psi_d},m_\phi)\; {\rm and}\; m_\phi < {\rm min}( m_{\psi_\chi},
m_\chi)$ so there are kinematically allowed decay channels of the scalar
$\chi$ into the $\psi_d,\bar{\psi}_d$ fermions. The decaying of the
inflaton field in this model is an indirect effect, 
interpreted in terms of the effective theory for $\phi$ (after
integrating over the $\chi, \psi_d, \bar{\psi}_d, \psi_\chi,
\bar{\psi}_\chi$ fields), which shows that the inflaton does not
interact for instance with vacuum like $\chi$ excitations but rather
with the collective $\chi$ excitations which can decay into light
fermions. These are scattering like processes through which the inflaton
transfers its energy (or radiates) and that can be efficient in 
nonlinear and nonperturbative regimes and can happen even deep inside the
inflationary phase, as shown
in [~\refcite{br4}]. 

The role of the spinors coupled to $\phi$ in (\ref{Lint}) is to mimic
SUSY, keeping the quantum corrections $\Delta V_{\rm eff} (\phi)$ to the
effective potential for the inflaton under control, so preserving the
flatness of its potential. In fact, as shown in
[~\refcite{br3}], a minimal SUSY model that reproduces the above
interactions and decay mechanism has the superpotential $W=
\sqrt{\lambda} \Phi^3 + g \Phi X^2 + f X^3 + m X^2 + h X Y^2$, where
$\Phi, X,Y$ are chiral superfields. Here, even for
SUSY breaking, for $\phi \neq 0$ and $H \neq 0$, we still
can have $\Delta
V_{\rm eff} (\phi) \ll V_0(\phi)$, where $V_0(\phi)$ is the tree level
potential for the inflaton, that we will take to be a quartic potential with
self-coupling $\lambda$.

The effective equation of motion (EOM) that emerges for a homogeneous
classical inflaton field, $\phi \equiv \varphi(t)$, from (\ref{Lint})
and with parameters satisfying the adiabatic (or slow) dynamics for the
inflaton, $\lambda \sim O(10^{-13}),\; g\sim g'\sim h \;\gtrsim\;
O(10^{-1})\; $ and $\; \varphi \sim O(m_{\rm Pl})$, was shown in Refs.
[~\refcite{br2,br4}] to be given by

\begin{eqnarray}
{\ddot \varphi}(t) \! +\!  3H(t) {\dot \varphi}(t) \!+\! 
\frac{dV_{\rm eff}(\varphi)}{d\varphi} \!+\! \xi \varphi(t) R(t)
\!+ \! 4 g^4 \varphi(t)  \!\! \int_{t_0}^t \!\!
dt' \! \varphi(t') \dot{\varphi}(t') K_{\chi} (t,t'),
\label{eom}
\end{eqnarray}  

\noindent 
where we are working in a FRW background metric, $R(t)$ is the
scalar of curvature, with coupling $\xi$ of the inflaton to the gravitacional
field and $K_{\chi} (t,t')$ is a nonlocal
(dissipative) kernel that results from the interaction of the inflaton
with the scalar $\chi$ within the relevant scattering like term at one-loop order,

\begin{eqnarray}
K_{\chi}(t,t') \!=\! \! \int_{t_0}^{t'} \!\! \frac{d\tau }
{a(t)^{3}} \! \int \! \frac{d^3 {\bf q}}{(2 \pi)^3}  
\sin\left[2\int_{\tau}^t \!\! dt'' \omega_\chi({\bf q},t'') \right] \;
\! \frac{ e^{-2 \int_{\tau}^t dt''\Gamma_\chi({\bf q}, t'')}  }
{ 4 \omega_\chi({\bf q},t) \omega_\chi({\bf q},t')} \Bigr|_{t>t'}
\label{kernel}
\end{eqnarray}

\noindent 
where $\omega_\chi ({\bf q},t) =
\left[{\bf q}^2/a(t)^2 + M_\chi^2(t) \right]^{1/2}$, $M_\chi^2(t) =
m_{\chi}^2 + g_j^2 \varphi(t)^2 + (\xi-1/6) R(t)$ and
$\Gamma_{\chi}({\bf q}, t) \simeq h^2 M_\chi^2(t)/\left[
8\pi\omega_\chi({\bf q},t)\right]$ for $m_\chi \gg m_{\psi_d}$. 

In Refs. [~\refcite{br3,br4}] we have studied the dynamics of the
inflaton through the full numerical solution of (\ref{eom}) (which is
numerically implementable, since $\Gamma_\chi > H$ and so the
highly oscillatory nonlocal kernel is effectively damped). We also have 
shown that
approximating the kernel as a nonexpanding one (in the Minkowski
approximation of Refs. [~\refcite{br1,br2}]) is a very good
approximation for the exact numerical dynamics as well (which is expected since
for our parameters $M_\chi \gg H$ and so curvature effects are
subleading). {}Finally we also have shown that a Markovian (local)
approximation for (\ref{eom}) is as well an excellent approximation to
describe the evolution for the inflaton (which again is expected, since
during inflation and parameters we consider, 
$\dot{\varphi}/\varphi, H < \Gamma_\chi$ and so the
dynamics is effectively adiabatic). 

A representative example of the effects of dissipation in the
inflaton's EOM as a result of its interactions to other fields, in
the relevant region of parameters for our mechanism of dissipation to
work, is shown in 
{}Fig. 1.

\begin{figure}[ht]
\epsfxsize=8cm   
\centerline{\epsfbox{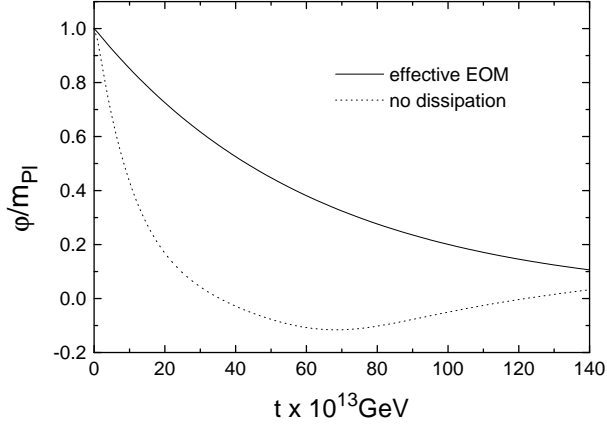}}
\caption{Evolution for $\varphi(t)$ for  
$g=h=0.5$, $\xi=0$,
$\lambda=10^{-13}$, $m_\chi= 10^{13}$GeV, 
$\varphi(0)=m_{\rm Pl}$, $\dot{\varphi}(0)=0$ and $a(0)=1$. }
\end{figure}

{}Figure 1 is obtained by numerically solving Eq. (\ref{eom}) in the Markovian
approximation simultaneously with the acceleration equation for the scale
factor. {}For comparison we also show the
result for $\varphi(t)$ when the dissipation due to the nonlocal term in
Eq. (\ref{eom}) is absent, for the case of a quartic effective
potential, $V_{\rm eff}= \lambda \varphi^4/4$ (dotted line). In the
absence of the nonlocal term in Eq. (\ref{eom}), inflation
for the quartic potential with the parameters of Fig. 1 ends when
$\varphi_{\rm end} \sim 0.47 m_{\rm Pl}$, or by the time $t_{\rm end} \sim
10^{-12}$ GeV$^{-1}$. At this time and well after the inflaton starts
oscillating around its minimum value of the potential, in the presence of
the effective dissipation term in Eq. (\ref{eom}) the inflaton is still
in the inflationary regime (the solid curve in Fig. 1), which ends by
the time $\sim 1.1 \times 10^{-11}$ GeV$^{-1}$, when $\varphi_{\rm end}
\sim 0.17 m_{\rm Pl}$. Till the end of inflation the dynamical regime
for the inflaton is overdamped, dominated by the nonlocal dissipative
kernel. Despite the noticeable change in behavior due to field
dissipation, the overall amount of radiation energy density produced 
is only a fraction of the inflaton's energy density.
For the parameters of {}Fig. 1, the radiation energy density reaches a
peak value $\rho_r/\rho_\varphi
\sim 10^{-2}$ at an early time, decaying next till reaching an approximate
constant fraction value of $\sim 10^{-5}$. We have also checked that the
adiabatic approximation used to derive Eq. (\ref{eom}),
$\dot{\omega}_\chi/\omega_\chi^2 \ll 1$ and the Markovian approximation
for the nonlocal kernel, $\dot{\varphi}/(\varphi \Gamma_\chi) \ll 1$
(see Ref. [~\refcite{br4}]), are both very robust, breaking down, for
the parameters used in Fig. 1, at a time $\sim 3 \times 10^{-11}$
GeV$^{-1}$ and therefore well after the end of inflation.

\begin{figure}[ht]
\epsfxsize=9cm   
\centerline{\epsfbox{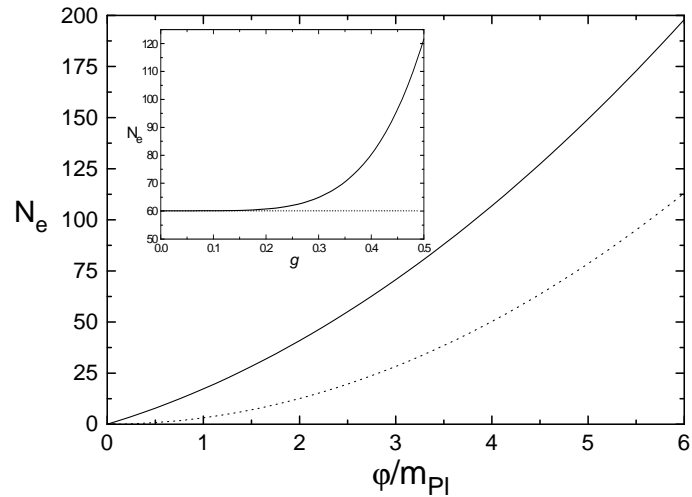}}
\caption{The number of e-folds $N_e$ in terms of $\varphi(0)$, 
considering the effective evolution (solid
line) and without dissipation (dotted line). Parameters are the
same as in Fig. 1. Inner plot shows $N_e$ as a function of
the coupling constant, with fixed $\xi=0$,
$\lambda=10^{-13}$, $m_\chi= 10^{13}$GeV,  $\varphi(0)=4.4 m_{\rm Pl}$,
$\dot{\varphi}(0)=0$. }
\end{figure}

{}Figure 2 compares how the number of e-folds of inflation, $N_e$,
changes when we vary either the initial inflaton's amplitude or
couplings $g,h$ for the interactions terms in Eq. (\ref{Lint}) and again
contrast the results with those obtained for the quartic potential for
the inflaton, in the absence of the interactions $L_{\rm int}$. {}For 
$\varphi(0)=4.4 m_{\rm
Pl}$, which for the inflaton's self-interaction $\lambda = 10^{-13}$
results in $N_e\simeq 60$ in the absence of dissipation effects in the
inflaton's EOM (dotted line in both plots shown in Fig. 2), the inner plot
in Fig. 2 shows the number of e-folds when we vary $g$ (taking also $h
=g$). We observe that the interaction terms (\ref{Lint}) start to
influence the inflaton's evolution in an appreciable way for $g=h \gtrsim
0.2$, with number of e-folds quickly raising up as the couplings are
increased. Increasing the number of fields $\chi$ or fermions $\psi_d$
has also similar effect of increasing rather quickly the number of e-folds
or the duration of the inflationary phase.

The results discussed above show that in typical multi-field inflation
models there are parameter regions feasible to inflation for which nonlinear and
nonperturbative effects can become important and that can lead to
important changes in the dynamics for the inflaton, with the
emergence of effective strong dissipative effects that alone can sustain
inflation long enough and with observational effects on density
perturbations \cite{br3,br4}. We should note that the appearance of
strong dissipative effects in our mechanism is not related to a direct
decay for the inflaton field, but it is a consequence of decaying modes
for fields coupled to the inflaton that results in an effective
dissipation in the inflaton's EOM, whose magnitude can be expressive for
nonlinear and nonperturbative regimes. These dissipative mechanisms
discussed here have found several uses in the recent literature, like in
alleviating many of the problems associated with typical inflation
models (the $\eta$ problem, graceful exit, quantum-to-classical
transition, large inflaton amplitude, initial conditions), in the study
of baryogenesis during nonisentropic inflation, among other studies.

\end{document}